\documentclass[%
 reprint,
superscriptaddress,
 amsmath,amssymb,
 aps,
]{revtex4-1}

\usepackage{placeins}
\usepackage{dcolumn}
\usepackage{graphicx}
\usepackage{bm}
\usepackage{physics}
\usepackage{import}
\usepackage{float}
\usepackage{amsmath}
\usepackage{amssymb}
\usepackage{transparent}
\usepackage{wrapfig}

\begin{document}
\title{Symmetry effects on the static and dynamic properties of coupled magnetic oscillators}
\author{J. P. Patchett}
\affiliation{Cavendish Laboratory, University of Cambridge, Cambridge CB3 0HE, United Kingdom} 
\author{M. Drouhin}
\affiliation{Hitachi Cambridge Laboratory, Cambridge CB3 0HE, United Kingdom} 
\author{J. W. Liao}
\affiliation{Cavendish Laboratory, University of Cambridge, Cambridge CB3 0HE, United Kingdom} 
\author{Z. Soban}
\affiliation{Institute of Physics, Academy of Science of the Czech Republic, Cukrovarnická 10,
	162 00 Praha 6, Czech Republic}
\author{D. Petit}
\affiliation{Cavendish Laboratory, University of Cambridge, Cambridge CB3 0HE, United Kingdom}
\author{J. Haigh}
\affiliation{Hitachi Cambridge Laboratory, Cambridge CB3 0HE, United Kingdom} 
\author{P. Roy}
\affiliation{Hitachi Cambridge Laboratory, Cambridge CB3 0HE, United Kingdom} 
\author{J. Wunderlich}
\affiliation{Universität Regensburg, Universitätstraße 31, 93040 Regensburg, Germany}
\affiliation{Institute of Physics, Academy of Science of the Czech Republic, Cukrovarnická 10,
	162 00 Praha 6, Czech Republic} 
\author{R. P. Cowburn}
\affiliation{Cavendish Laboratory, University of Cambridge, Cambridge CB3 0HE, United Kingdom} 
\author{C. Ciccarelli}
\affiliation{Cavendish Laboratory, University of Cambridge, Cambridge CB3 0HE, United Kingdom}

\date{\today}
	\begin{abstract}
	 The effect of symmetry on the resonance spectra of antiferromagnetically coupled oscillators has attracted new interest with the discovery of symmetry-breaking induced anti-crossings. Here, we experimentally characterise the resonance spectrum of a synthetic antiferromagnet Pt/CoFeB/Ru/CoFeB/Pt, where we are able to independently tune the effective magnetisation of the two coupled magnets.  To model our results we apply the mathematical methods of group theory to the solutions of the Landau Lifshitz Gilbert equation.
  This general approach, usually applied to quantum mechanical systems, allows us to identify the main features of the resonance spectrum in terms of symmetry breaking and to make a direct comparison with crystal antiferromagnets. 
  
	\end{abstract}
\maketitle

\section{Introduction}

The analysis of symmetry is important across all sub-fields of physics, and has recently been employed to the analysis of coupled magnetic oscillators (CMO). A quantum mechanical picture based on group theory has been applied to ferrimagnets such as Yttrium-Iron-Garnet (YIG) to calculate the dispersion of the modes that are deep into the Brillouin zone \cite{CHEREPANOV199381}. 
However, the majority of experimental investigations on CMOs employ lab-bench ferromagnetic resonance techniques, which are capable of exciting and detecting modes in the frequency range $\leq 100$ GHz with wave-vectors close to the centre of the Brillouin zone. Within these experimental conditions, a theoretical description in terms of quantum mechanics might appear un-necessary and cumbersome. Instead, most experimental results are analysed using the classical Landau-Lifshitz-Gilbert (LLG) equation \cite{Landau, Gilbert1955}. Recently, the implications of symmetry and symmetry-breaking on the solutions of the LLG equation have been explored via a theoretical description of magnon-magnon anti-crossings in anti-ferromagnetically coupled oscillators. In all these works the breaking of symmetry, either externally by applying a magnetic field or internally by tuning the material composition, was identified as  key in preventing two modes' crossing \cite{Liensberger2019, Macneill2019, PhysRevLett.125.017203, Liu2014, Zhang1994,Dutra2013}. In these reports, the results are analysed mathematically by looking at the commutation of various transformation matrices and the resulting conserved quantities. By re-expressing the condition for crossing/anti-crossing in terms of group theory, the usual mathematical technique for analysing symmetry in other systems of coupled oscillators, it is possible to not only analyse crossings from a symmetry perspective, but also other features of a magnetic resonance spectrum from a symmetry perspective, such as symmetry induced degeneracies of the modes.

In this work we apply general symmetry arguments based on group theory to elucidate the mechanisms of magnon mode splitting. We measure the ferromagnetic resonance (FMR) spectrum of a Pt/CoFeB/Ru/CoFeB/Pt synthetic antiferromagnet (SAF). Here, we are able to engineer the level of asymmetry between the two coupled magnetic layers by independently tuning the out of plane anisotropy $H_{a}^{N}$ of the two magnets, thus the effective magnetisation $M_{s}^{eff}= M_{s}-H_{a}^{N}$, via the thickness of Pt.
By showing that the solutions of the LLG equation engender a representation of the symmetry group of the system, with minimal modification to the familiar case of Hermitian eigenvalue problem,  we interpret the symmetry induced anti-crossing as a bare manifestation of the Wigner Von-Neuman anti-crossing theorem \cite{WIGNER}. Furthermore, by extending our analysis beyond the anti-crossings we are able to interpret other features of the resonance spectra, in particular the degeneracy (or lack of it) of the modes at zero field. 

Finally, we apply a separate analysis of the magnetoresistance to confirm the degree of asymmetry between the coupled magnets.

We begin by the application of group theory to the solutions of the LLG equation for a system of antiferromagnetically coupled oscillators, we follow on with an analysis of the effect of symmetry on the equilibrium magnetic configurations in section 3. Finally, we apply these results to analyse our experimental measurements in section 4.

\begin{figure}
	\centering
\includegraphics[width=1.0\linewidth]{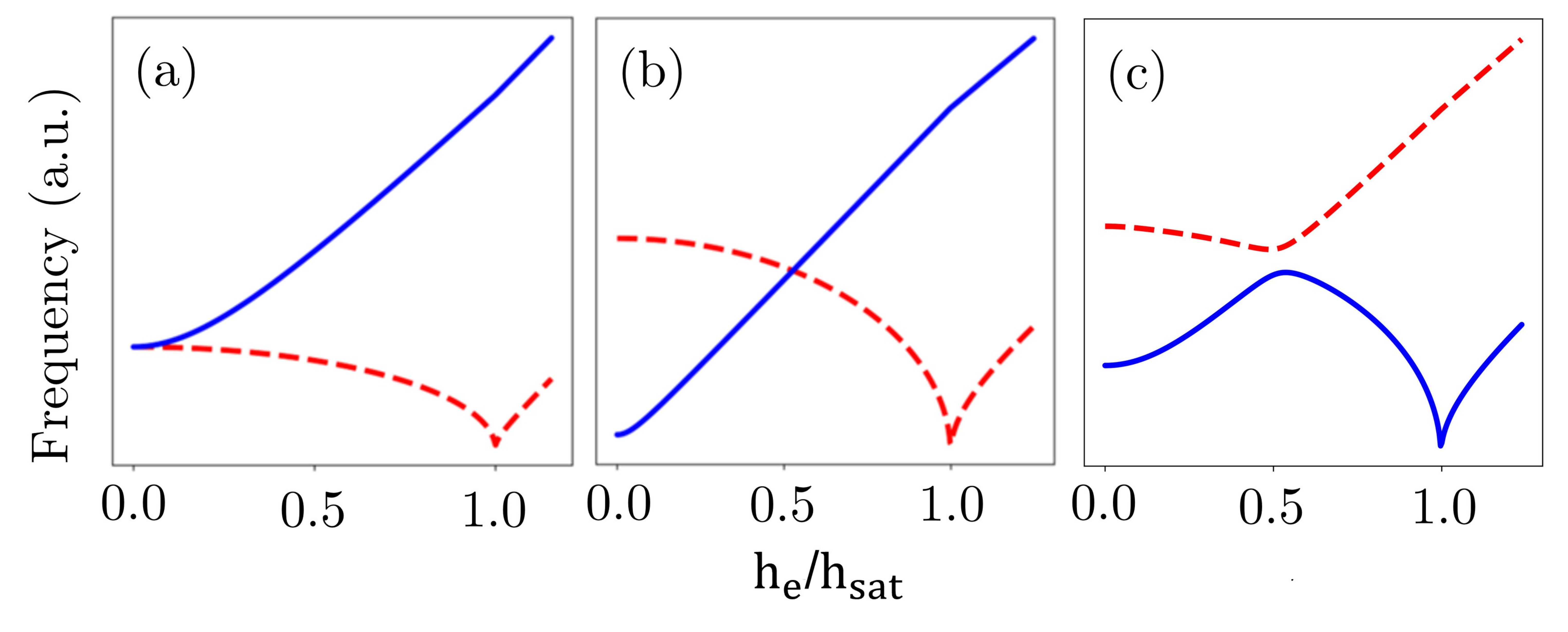}
	\caption{Resonance frequencies vs field of a uniaxial antiferromagnet (a) and a synthetic antiferromagnetic with symmetric (b) and asymmetric (c) components when the external magnetic field is applied along the hard axis.}
	\label{fig:sffmodes}
\end{figure}

\section{Dynamic theory}

\begin{figure}
	\centering
\includegraphics[width=1.0\linewidth]{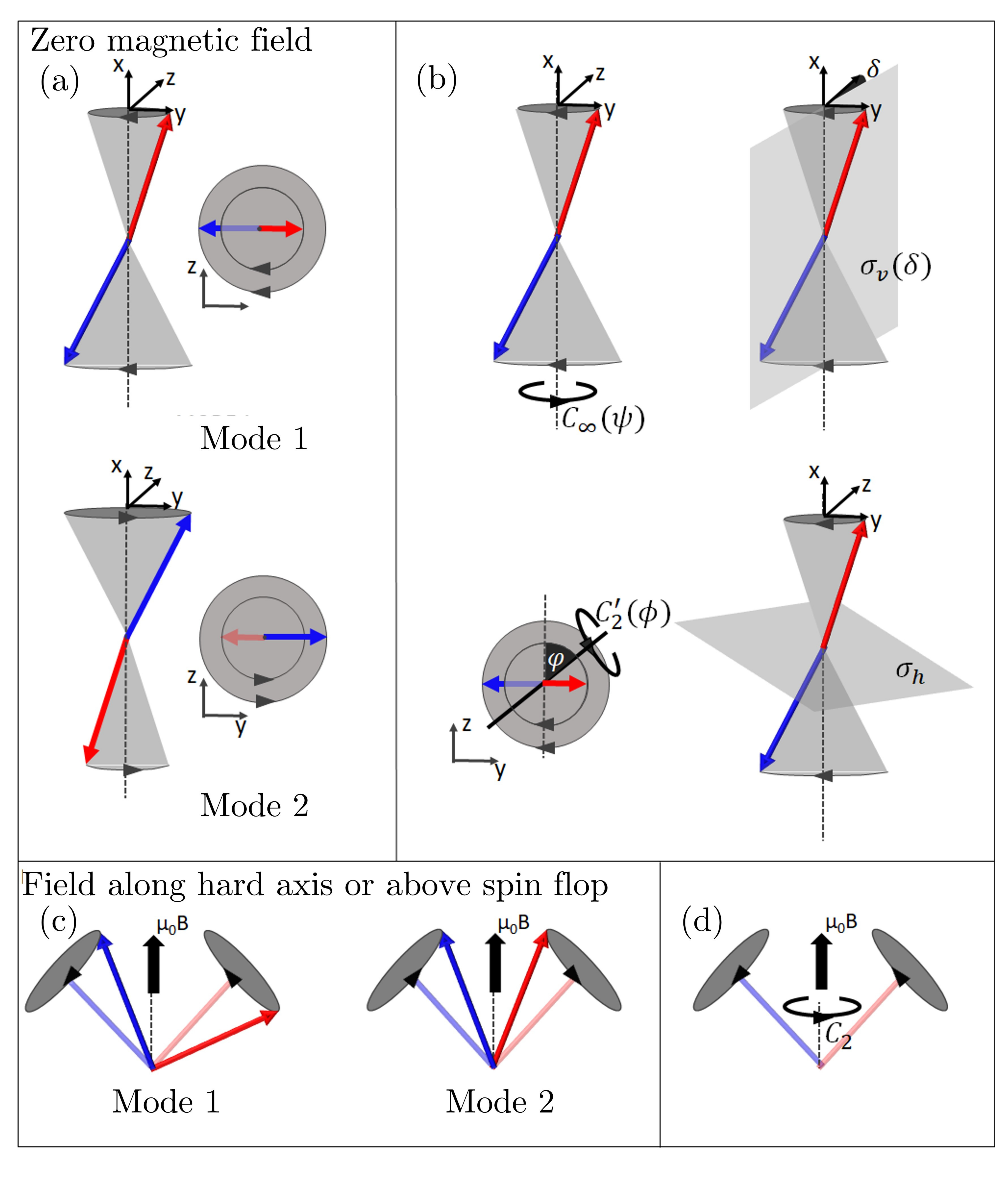}
\caption{Pictorial representation of the resonance modes (a) and some of the symmetry operations (b) of a uniaxial antiferromagnet at zero magnetic field.  Resonance modes (c) and symmetry operation $C_{2}$ (d) for a system of thin-film antiferromagnetically coupled oscillators when the magnetic field is applied along the hard axis or is above the spin flop field.}
\label{fig:zeroFM}
\end{figure}

We model the oscillators by using a macrospin approximation \cite{KITTEL1948} for the two coupled magnetic layers, described by the normalised magnetisation vectors $\bm{m}_{A}$ and $\bm{m}_{B}$. The LLG equation \cite{Landau} describes the dynamics of $\bm{m}_{A/B}$. Starting from the expression of the linearised LLG equation by Smit and Beljers \cite{SmithandBeljers} in the dissipationless limit, the LLG equation can be transformed into a generalised Hermitian eigenvalue equation for the resonance frequencies, $\omega_{\mu}$, and vectors $\bm{\mu}$ (please refer to section 4 of the supplementary information for a full description of the dynamic theory) \cite{Zivieri2012, PhysRevB.70.054409, PhysRevB.75.024416}:
\begin{equation}\label{eq:linearisedLLG}
\begin{split}
\omega_{\mu}\mathcal{I}\bm{\mu} = \gamma H\bm{\mu}\\
&\mathcal{I} = \mu_{0}\begin{pmatrix}
i\underline{A}(M_{sA}V_{A}) && 0 \\
0 && i\underline{A}(M_{sB}V_{B})
\end{pmatrix}~ ; ~ \underline{A}(x) = 
\begin{pmatrix}
0 && x \\
-x&& 0 
\end{pmatrix}
\end{split}
\end{equation} 
where $\gamma$ is the gyromagnetic ratio, $M_{si}$ and $V_{i}$ are the saturation magnetisation and volume of layer $i$, $\bm{\mu}$ is a 4-component vector describing the transverse oscillations of each layer's magnetisation, $H$ is the Hessian of the energy as a function of the normalised magnetisation and $\mathcal{I}$ is a matrix that originates from the cross product of the linearised effective field with the equilibrium magnetisation. This can be extended to an arbitrary number of coupled oscillators. The underlying Hermitian structure means that the modes form a complete basis and, with appropriate normalisation, are orthonormal under inner product with $\mathcal{I}$, i.e. $\bm{\mu}^{\dagger}\mathcal{I}\bm{\nu} = \delta_{\mu \nu}$. Consequently, they engender a representation of the symmetry group whose elements are the operations that leave both the device and magnetisation unchanged (we direct the reader to section 5 of the supplementary information for an introduction to group theory). Here, a symmetry operation $R$ is mapped to a matrix with components $R_{\mu \nu} = \det{R}\bm{\mu}^{\dagger}\mathcal{I}R\bm{\nu}$, where the additional $\det{R}$ factor accounts for the pseudo-vector nature of the magnetisation.

Coupled oscillators have two unique modes, which we refer to as Mode 1 and Mode 2 \cite{doi:10.1063/1.3225608, 6028142, Li2018}. We will first consider the case where no external magnetic field is applied. In a uniaxial antiferromagnet the two modes, shown in Fig.\ref{fig:zeroFM}(a), are degenerate at zero field (Figure \ref{fig:sffmodes} (a)). In a SAF, where the two magnets are thin films, the spectrum looks radically different with the degeneracy being removed (Figure \ref{fig:sffmodes} (b) and (c)). This difference can be understood in terms of the different symmetries that characterise the two types of antiferromagnetically coupled oscillators.

For a uniaxial antiferromagnet, the symmetry operations, i.e. the operations that leave the static configuration of the system invariant, are the same as those of the familiar homonuclear diatomic molecule and form the symmetry group $D_{\infty h}$. A selection of these symmetry operations is shown in Figure \ref{fig:zeroFM}(b) and their action on the two modes is summarised in table \ref{tab:Reps}.

\begin{table}[h]
	\centering
\begin{tabular}{c|c|c|c|c}
	$\bm{D}_{\infty h}$ & $C_{\infty}(\psi)$ & $\sigma_{v}(\delta)$   & $C_{2}'(\phi)$  & $\sigma_{h}$  \\ 
	\hline 
	$E_{1g}$ & $\begin{pmatrix}
e^{-i\psi} && 0 \\
-0 && e^{i \psi} 
\end{pmatrix}$ & $\begin{pmatrix} 0 && -e^{-i2\delta} \\ -e^{i 2 \delta} && 0 \end{pmatrix}$ & $\begin{pmatrix} 0 && e^{-i2\phi} \\ e^{i2\phi} && 0 \end{pmatrix} $ &  $\begin{pmatrix} -1 && 0 \\ 0 && -1 \end{pmatrix}$  \\ 
\end{tabular} 
\caption{ Representations of symmetry operations engendered by the degenerate eigenmodes of a uniaxial antiferromagnet at zero field.\label{tab:Reps}}
\end{table}

$C_{\infty}(\psi)$ describes an anticlockwise rotation about the easy axis by an angle $\psi$ and adds a phase $e^{-i\psi}$ and $e^{i\psi}$ to mode 1 and mode 2 respectively. $\sigma_{v}(\delta)$ is a reflection in a plane that contains the easy-axis and makes an angle $\delta$ with the z-axis. Following the usual rules for pseudo-vectors under reflection/inversion, this operation transforms the modes into one another with the additional phase $e^{-i2\delta}$ and $e^{i2\delta}$. $C_{2}'(\phi)$ is a 180 degrees rotation about an axis lying in the yz-plane at an angle $\phi$ with the z-axis and again transforms the modes into each other with the addition of a phase. $\sigma_{h}$, a reflection in the yz-plane perpendicular to the easy-axis, transforms the modes into themselves with the addition of a 180 degrees phase. The additional elements of $D_{\infty h}$, consisting of an inversion operator and improper rotations, can be calculated similarly. Together, the matrix representations engendered by the modes form the irreducible representation $E_{1g}$ of the full symmetry group $D_{\infty h}$ (please refer to section 5 of the supplementary information). A representation is said to be irreducible when we cannot identify any change in basis (basically two new modes, linear combination of the original ones) in which all the matrices that make the group are reduced in a block-diagonal form (or in the case of our 2-dimensional representation, diagonal form). In a 2-dimensional representation, being irreducible is therefore equivalent to having non-zero commutators for at least one pair of matrices. The matrices in Table \ref{tab:Reps} have non-zero commutators for $\psi \neq n \pi$ and $\phi$ and $\delta \neq n\pi/2$ and this means that they cannot be simultaneously diagonalised by a similarity transformation. For a 2D representation this is equivalent to the representation being irreducible and it results in the modes being degenerate at zero field. 

To lift the degeneracy it is necessary to lower the symmetry of the system such that the 2D irreducible representation of the higher symmetry group decomposes into two 1D irreducible representations of the new lower symmetry group. Equivalently, the symmetry operations with non-vanishing commutators must be removed from the group.
In a SAF, where the coupled magnets are thin films in the xy-plane and assumed to be equal, the presence of shape anisotropy leads to $C_{\infty}(\psi)$, $\sigma_{v}(\delta)$ and $C_{2}'(\phi)$ being symmetry operations only when $\psi = n\pi$ and $\phi$ and $\delta = n\pi/2$, which induces the splitting of the modes at zero field. Therefore, we can understand the presence or absence of degeneracy at zero field as a consequence of symmetry lowering.  

The implications of symmetry on the resonance spectra of SAFs has recently been discussed relatively to the anti-crossings observed between the acoustic and optical modes when an external field is applied. In these previous works the anti-crossing has been explained with the lowering of the system's symmetry either by applying a magnetic field away from an axis of high symmetry or by tuning the crystalline anisotropy \cite{Liensberger2019}, shape anisotropy \cite{Macneill2019, PhysRevB.102.100403}, or spin wave propagation direction for $k \neq 0$ magnons \cite{PhysRevLett.125.017203}. These results have been understood by deriving a conserved quantity from the commutation of the matrix describing the linearised LLG problem with a matrix representing the exchange of the two oscillators. In other disciplines, such as in the study of the vibrational modes of complex molecules or electronic energy levels in crystals, anti-crossing is commonly understood via the Wigner-Von Neuman anti-crossing theorem which states that only modes that transform according to different irreducible representations of their symmetry group may cross. It is straightforward to show that this still holds for coupled magnetic oscillators. Here we consider the case of a magnetic field applied along the hard axis direction or along the easy axis above the spin-flop value. The two modes of a symmetric SAF are shown in Fig.\ref{fig:zeroFM}(c). The resonance spectra of mode 1 and mode 2 correspond respectively to the continuous blue and dashed red lines in Fig.\ref{fig:sffmodes}(b). 

In a symmetric SAF where the two magnetic layers are equal, $C_{2}$, describing 180 degree rotations about the field direction (Figure \ref{fig:zeroFM}(d)), is a symmetry operation. The same is true for the field applied along the easy axis above the spin-flop value. The two modes thus transform as the irreducible representations of the group $A_{2}$ and $A_{1}$, as follows 

\begin{equation}
\centering
\begin{tabular}{c|c|c}
& $E$ & $C_{2}$ \\ 
\hline 
$A_{1}$ & 1 & 1 \\ 
\hline 
$A_{2}$ & 1 & -1 \\ 
\end{tabular}. \label{eq:characterTable}
\end{equation}

here $E$ is the identity transformation. Because these two modes transform according to different representations, from the Wigner von-Neumann avoided crossing theorem \cite{WIGNER}, they are predicted to cross at a single external field value. However, when the symmetry of the system is lowered, for example by making the two magnetic layers unequal or by applying the magnetic field in a direction that does not coincide with a high symmetry direction, $C_{2}$ is no longer a symmetry operation and the modes must transform according to the same trivial representation, causing them to hybridise strongly and repel (Fig.\ref{fig:sffmodes}(c)). \\

In what follows we will consider the static behaviour of antiferromagnetically coupled systems and provide an alternative method for verifying the asymmetry between the two coupled magnets. Brown's equation \cite{brown1963micromagnetics}, $\bm{M}_{(A/B)}\times \bm{H}^{(A/B)}_{eff} = \bm{0}$, determines the equilibrium orientation for each magnet. $\bm{M}_{i}$ and $\bm{H}_{eff}^{i}$ are the magnetisation (with saturation magnetisation $M_{si}$) and effective field in layer $i$. Within the macrospin approximation, the effective field can be written in the form:
\begin{equation}
\bm{H}^{i}_{eff} = \bm{\hat{\Lambda}}^{ij}\bm{M}_{j}+\bm{h}^{i}_{e}
\end{equation}
Where $\bm{\hat{\Lambda}}^{ij}$ is an operator that describes the field experienced by layer $i$ due to layer $j$. If the external field is transformed by an orthogonal operation $R$, $\bm{h}^{i}_{e}\rightarrow R\bm{h}^{i}_{e}$, where $R$ commutes with $\bm{\hat{\Lambda}}^{ij}$, then $R\bm{M}_{i}$ is a solution for the rotated field $R\bm{h}_{e}$ since $(R\bm{M}_{i})\times(R(\bm{\Lambda}^{ij}\bm{M}_{j}+\bm{h}^{i}_{e}) = \bm{0}$. Although the solutions of Brown's equation are by no means unique, if the entire hysteresis path, locus of the solutions $\bm{M}_{i}$ at each $\bm{h}^{i}_{e}$, is also transformed by $R$, then at $R\bm{h}^{i}_{e}$ the  magnetisation must be $R\bm{M}_{i}$ (we refer the reader to section 3 of the supplementary information for further details). 

In the special case where the in-plane easy axis is at $45^{\circ}$ or $135^{\circ}$ with respect to the current direction, a reflection in the plane perpendicular to the in-plane hard or easy axis induces a sign change in the anisotropic magnetoresistance (AMR) term of the magnetoresistance but leaves the giant magnetoresistance (GMR) term unchanged because the angle between the two magnets remains the same. Therefore, measuring the magnetoresistance for magnetic field sweeps along directions that are coupled by such reflection provides a way of discerning between the different magnetoresistance contributions in a SAF.

If the two coupled magnets are identical- both in terms of their magnetic and transport (resistance, AMR coefficient etc.) properties, this reflection is a symmetry operation when the magnetic field is applied perpendicular to the reflection plane, hence the magnetoresistance is unchanged. Consequently, when the AMR is the dominant magnetoresistive term, as we observe in our experiments described later, this must result in zero magnetoresistance thorough the field sweep.  In this way we have a method for establishing the \textit{degree of asymmetry} in the magnetotransport properties of the two magnets of the SAF. We note that the AMR coefficient in thin-films is effected by a number of interfacial effects, spin-dependent scattering, magnetic dead layers etc. In principle, it is possible, though unlikely, for these effects to cancel even in an asymmetric SAF, and consequently give a symmetric appearing AMR signal.

\section{Experimental results:}

\begin{figure}
	\centering
\includegraphics[width=1.0\linewidth]{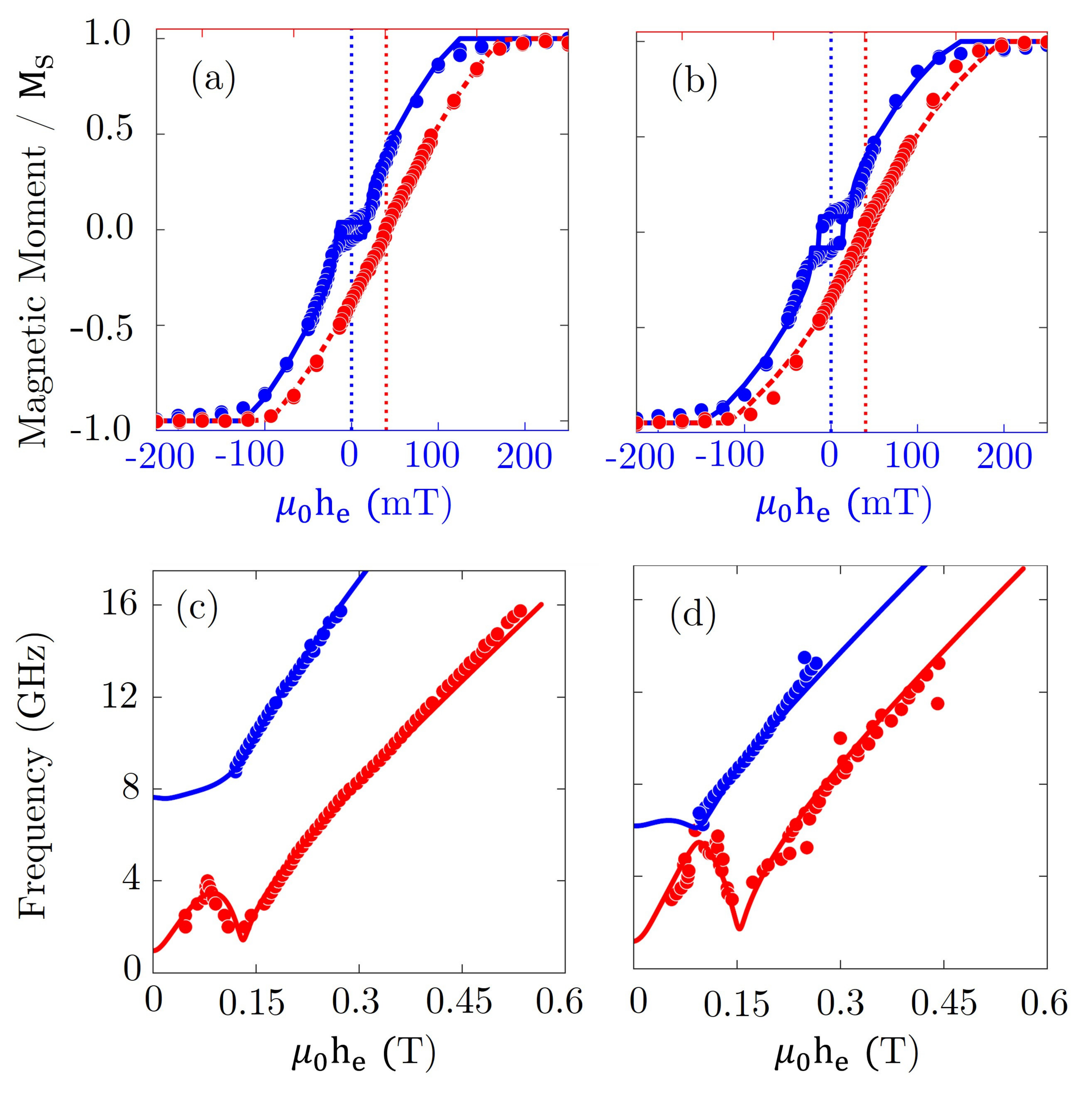}
	\caption{Net magnetic moment measured by VSM and FMR spectrum for a field applied along the in-plane hard axis for the single sided (a) \& (c) and double sided (b) \& (d) structures. In (a) and (b), the blue (red) data set represents the net magnetic moment when the external field is applied along the easy (hard) axis. Vertical dashed lines correspond to zero external field (a lateral shift between the two graphs is introduced for clarity purposes).}
	\label{fig:FMRspectrum}
\end{figure}

We measure two structures, Ta[2]/Pt[3]/CoFeB[1.4]/Ru[0.9]/CoFeB[1.4]/Pt[x]/Ta[2] (all thicknesses in nm) where x = 1 nm (referred to as the single sided structure) or 6 nm (referred to as the double sided structure). The SAF structures were fabricated using DC magnetron sputtering with base pressure $\sim 10^{-8} ~\mathrm{mbar}$ and an Ar pressure of $7\times 10^{-3}~\mathrm{mbar}$ during deposition.  An external magnetic field was applied during sputterring to induce an in-plane uni-axial anisotropy. An additional single magnetic layer structure Ta[2]/Pt[3]/CoFeB[1.4]/Ru[0.9]/Ta[2] was also fabricated under the same conditions and used to characterise the anisotropy (please refer to section 2 of the supplementary information). We extracted an effective magnetisation of $211 \pm 2 ~\mathrm{mT}$, significantly lower than the bulk magnetisation of CoFeB ($\approx 1500 ~\mathrm{mT}$), owing to the  significant out of plane anisotropy. 

\begin{figure}[h]
	\centering
	\includegraphics[width=1.0\linewidth]{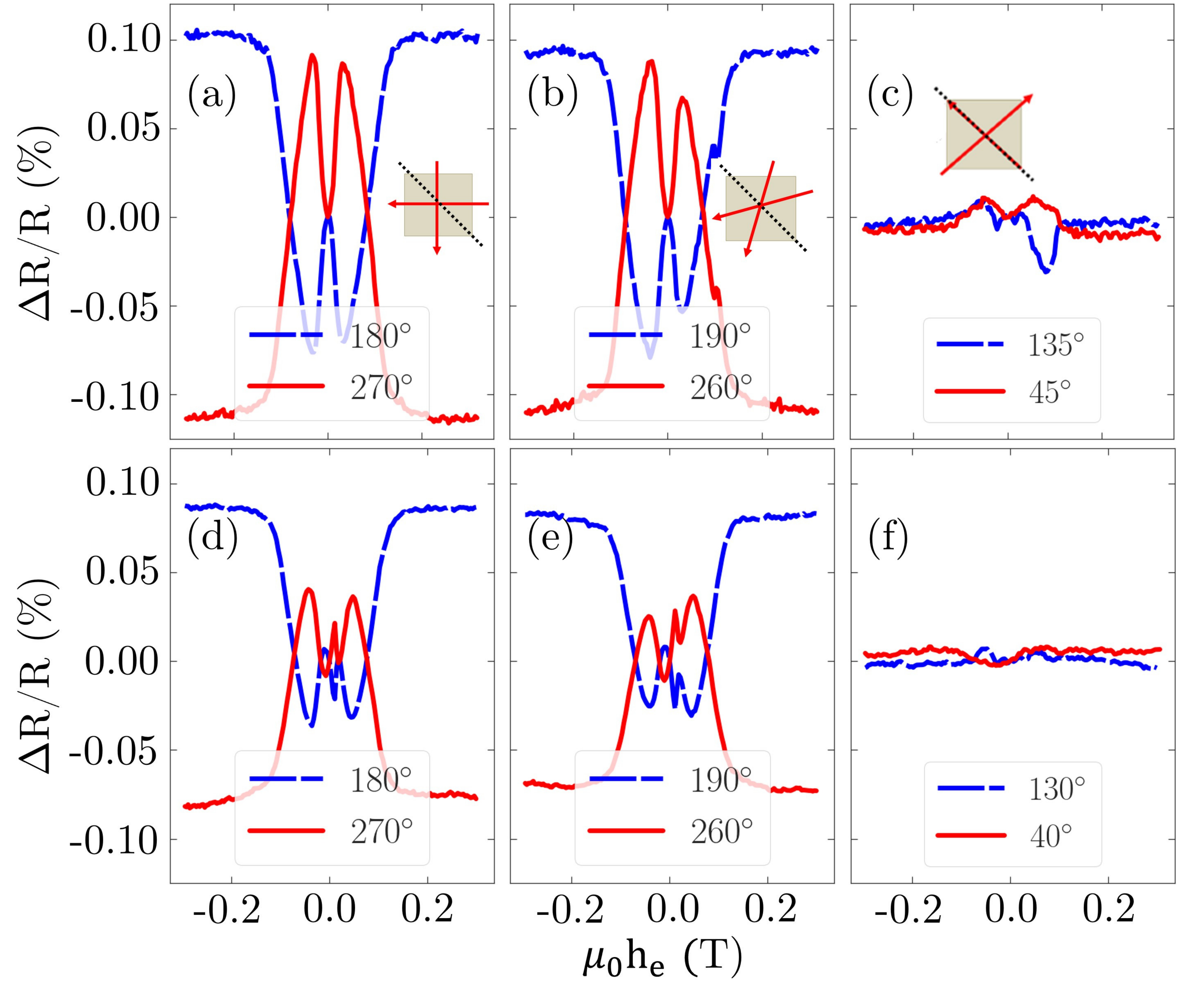}
	\caption{Magnetoresistance data for the single sided ((a) and (b)) and double sided ((d) and (e)) structures when the magnetic field is swept from negative to positive values along different directions connected by symmetry. Additionally, the magnetoresistance is measured along the hard and easy axes ((c) and (f)).  Inserts: cartoons showing the magnetic field sweep directions (red and blue arrows) relative to the easy axis (dotted line).}
	\label{fig:amrds}
\end{figure}

The devices were patterned into $5~\mathrm{\mu m}\times 5 ~\mathrm{\mu m}$ resistors by electron beam lithography and argon ion milling and electrically contacted on two opposite sides. The low aspect ratio for the resistor is chosen so as not to further lower the symmetry of the device by introducing additional shape anisotropy terms (the in-plane demagnetisation field is estimated to be $\sim 1 ~\mathrm{mT}$, negligible with respect to the other anisotropy terms). The orientation of the device is such that the in-plane field-induced easy axis cuts along one diagonal of the square, 135 degrees from the current flow direction.

\begin{table*}
\caption{Numerical parameters calculated from FMR fitting for layers A/B. \label{tab:FMRfit}}
\begin{ruledtabular}
\begin{tabular}{cccccc}
	Structure &  $\Delta(M_{s}^{eff})~(\%)$  & $\Delta (M_{s}t)~(\%)$ & $J^{(1)}~(\mu J/m^{3}) $  & $J^{(2)}~(\mu J/m^{3})$ &  $\mu_{0}H_{a}^{(P)}~(\mathrm{mT})$ \\ \hline
 	Single Sided &   $610 \pm 40$ & $7.5 \pm 0.6 $ & $50.5 \pm 0.5 $ &  $6.7 \pm 0.6$ & $3 \pm 2$  \\
	Double Sided &   $13 \pm 9$ & $15.2 \pm 1.2 $ & $58.6 \pm 0.6$  & $12.2 \pm  1.0$ & $2 \pm 2$ \\
\end{tabular}
\end{ruledtabular}
\end{table*}
To investigate the effects of symmetry-break between the layers on the dynamic properties of the synthetic antiferromagnets, we measure the magnetic field dependence of the resonance frequencies by current-induced FMR (please refer to section 1 of the supplementary information for further details on the measurement layout) and the net magnetic moment by vibrating sample magnetometer (VSM), as shown in Figure \ref{fig:FMRspectrum}. The two plots are fitted simultaneously (continuous line in the figure) via least squares regression using a custom macrospin simulation. The equilibrium position of the two magnetic moments for a certain value of the external magnetic field is found by numerically simulating the LLG equation via a finite difference method with $\alpha \neq 0$ until convergence. Then, the resonance frequencies are determined numerically by solving the eigenvalue equation \eqref{eq:linearisedLLG} using the free energy density given in section 4 of the supplementary information. The extracted parameters are summarised in Table 1. $\Delta(M_{s}^{eff})$ and $\Delta(M_{s}t)$ represent respectively the relative difference in effective and saturation magnetisation of the top magnetic layer relative to the bottom one, $J^{(1)}$ and $J^{(2)}$ are the bilinear and biquadratic exchange constants, while $\mu_0 H_{a}^{(P)}$ is the in-plane uni-axial anisotropy, assumed identical in both layers within experimental error. The double sided sample behaves in a similar manner to a perfectly symmetric SAF, with no experimentally observable anticrossing. The similar magnetic properties between the layers imply that $C_{2}$ remains a symmetry operation and no anticrossing is therefore measured. In the single sided structure we find that the out-of-plane anisotropy of the upper magnetic layer is significantly reduced compared to the lower layer, resulting in a large difference in effective magnetisation which removes $C_{2}$ as symmetry operation and results in significant anticrossing.

Finally, in Figures \ref{fig:amrds} (a), (b), (d) and (e) we show the change in longitudinal resistance measured as the field is swept along different in-plane directions. We see that the magnetoresistance of both devices is antisymmetric for specular directions of sweep with respect to the easy and hard axis, confirming that AMR is the dominant source of magnetoresistance in our synthetic antiferromagnetic structures. Figures \ref{fig:amrds} (c) and (f) show the AMR signal when the external field is applied along a high symmetry direction. While in the double sided sample (Figure \ref{fig:amrds} (f)) the magnetoresistance variations are within the error, in the single sided sample (Figure \ref{fig:amrds} (c)) we measure a non-zero magnetoresistive response, which is indicative of a difference in the magnetotransport properties of the two magnetic layers. Our method does not allow identifying the origin of this anisotropy. One possibility is in the different quality of the interfaces, which affects the nature of the scattering in the region closer to the interface and the current distribution among the different layers.

In conclusion, in this work we have applied a symmetry analysis based on group theory to study the resonance spectrum of a system of coupled magnetic oscillators. Our analysis captures the main differences in the spectrum of crystalline and synthetic antiferromagnets
and provides an alternative explanation of the symmetry protected crossing between the optical and acoustic modes in terms of the Wigner Von-Neuman theorem. Additionally, by employing group theory one can extend the symmetry analysis to look at the degeneracy of the resonance modes. We have applied this analysis to the specific case of a SAF where we were able to tune the level of asymmetry between the coupled magnets by varying the out-of plane anisotropy.    

\section*{Acknowledgements}
The authors thank Jakub Zelezny for the useful discussions and for checking the theory, and Thomas Wagner for his help with the ferromagnetic resonance measurements.
JPP acknowledges support from the EPSRC Doctoral Training Programme. WL acknowledges support from a Newton International Fellowship scheme NF150217. ZS is supported by the  Ministry of Education of Czech Republic, Grant No. LM2018110 and LNSM-LNSpin. J.W. acknowledges funding from the ERC Synergy Grant No. 610115, from Ministry of Education of the Czech Republic Grant No. 451 LM2018110 and LNSM-LNSpin, the Czech Science Foundation Grant No. 19-28375X, and the EU FET Open RIA Grant 453 No. 766566; CC acknowledges support from the Royal Society and the Winton Programme.

\FloatBarrier
\bibliographystyle{ieeetr}
\bibliography{main}

\end{document}